\documentclass{ws-ijmpa}
\newcommand{\hk}{\hat{k}}
\newcommand{\hP}{\hat{P}}
\newcommand{\M}{{\cal M}}
\newcommand{\eps}{\varepsilon}

\newcommand{\dels}{\delta q^s}
\newcommand{\delv}{\delta q^v}

\newcommand{\ai}{\tiny a_{1}}
\newcommand{\aiNN}{\scriptstyle a_{1}NN}

\newcommand{\bia}{\scriptstyle b_{1}}
\newcommand{\biNN}{\scriptstyle  b_{1}NN}
\newcommand{\hi}{\scriptstyle h_{1}}
\newcommand{\hiNN}{\scriptstyle  h_{1}NN}
\newcommand{\hit}{h_{1}(1170)}
\newcommand{\hitNN}{\scriptstyle h_{1}(1170)NN}
\newcommand{\gmnn}{g_{\mbox{$\scriptstyle {}_{\M N N}$}}}

\newcommand{\nnn}{\nonumber }

\newcommand{\be}{\begin{eqnarray}}
\newcommand{\ba}{\begin{array}}
\newcommand{\ea}{\end{array}}
\newcommand{\ee}{\end{eqnarray}}

\newfont{\fib}{cmfi10 at 10pt}

\newcommand{\Tr}{{\rm Tr}}

\newcommand{\da}{\dagger}

\begin{document}
%

\def\nocropmarks{\vskip5pt\phantom{cropmarks}}


%

\markboth{L. Gamberg and G.R. Goldstein}
{Flavor-Spin Symmetry and the Tensor Charge}

%
\catchline{}{}{}
%

\setcounter{page}{1}

\title{FLAVOR-SPIN SYMMETRY AND THE TENSOR CHARGE\footnote{Talk given by L.G. 
at $3^{\rm rd}$ Circum Pan-Pacific Symposium on High Energy Spin Physics, 
Beijing, China 8-13, Oct. 2001. 
}}

\author{\footnotesize LEONARD GAMBERG\footnote{
gamberg@dept.physics.upenn.edu}}

\address{Department of Physics and Astronomy, DRL 209 S. 
$33^{\rm rd}$ University of Pennsylvania,\\
Philadelphia, PA 19104-6396\ \ USA
}

\author{GARY R. GOLDSTEIN\footnote{ggoldst@tufts.edu}}

\address{Department of Physics and Astronomy, Tufts University, \\Medford, MA
 02155\ \ USA
}

\maketitle


\begin{abstract}
Exploiting  an approximate phenomenological  symmetry of 
the $J^{PC}=1^{+-}$ light axial vector mesons 
and using pole dominance, we calculate the flavor contributions
to the nucleon tensor  charge. 
The result depends on the decay constants of the 
axial vector mesons and their couplings to the nucleons.
\end{abstract}

\section{Introduction}

Investigations of the spin composition of the nucleon have led to 
surprising insights, beginning with the
revelation that the majority of its spin is carried by quark and
gluonic orbital angular momenta and gluon spin rather than by quark
helicity.\cite{smc,ji_spin}  In addition, considerable effort has gone into
understanding, predicting and measuring the
transversity distribution, $h_1(x)$, of the nucleon.\cite{review}
{\it Transversity}, as combinations of helicity 
states, $|\bot /\top> \sim \left(\, |+> \pm |->\, \right)$, for 
the moving nucleon is a variable introduced originally by Moravcsik and
Goldstein\cite{transversity} to reveal an underlying simplicity in
nucleon--nucleon spin dependent scattering amplitudes.
In their analysis of the chiral odd
distributions, Jaffe and Ji\cite{jaffe91} related the first moment of
the transversity distribution to the flavor contributions
of the nucleon tensor charge:
$\int_0^1 \left(\delta q^a(x)-\delta\overline{q}^a(x)\right) dx=\delta q^a$
(for flavor index $a$).  The leading twist 
transversity distribution function, $\delta q^a(x)$,
is as fundamental to understanding the spin structure of the nucleon
as  its helicity counterpart $\Delta q^a(x)$.
While the latter in principle can be measured in hard 
scattering processes, the transversity distribution 
(and thus the tensor charge)  decouples at leading twist in deep
inelastic scattering since it is chiral odd. 
Bounds placed on the leading twist quark 
distributions through positivity constraints suggest
that they  satisfy the inequality of 
Soffer,\cite{soffer95}
yet, the non-conservation of the tensor charge  makes it difficult
to predict.
In contrast to the axial vector isovector charge, no sum rule
has been written that enables a clear relation between the tensor charge
and a low energy measurable quantity. 
Among the various approaches, 
from the QCD sum rule to lattice calculation
models,\cite{jihe} there appears 
to be a range of expectations and a disagreement 
concerning the sign of the down quark contribution.

Recently, further  insight into tranversity 
has come from its interpretation in terms of Deeply Virtual Compton
Scattering (DVCS) 
in the context of skewed parton distributions (SPDs)\cite{rad} wherein
those distributions which flip the quark
helicity are the skewed counterparts of the usual
quark transversity distributions.  
In the forward limit  $H^a_T(x,\xi,t)$ reduces to
the ordinary tranversity distribution, $H^a_T(x,0,0)=\delta q^a(x)$.
It's first moment is nothing other than
the $t\rightarrow 0$ limit of the form factor 
associated with the quark  helicty flip amplitude $A_{++,--}$ which
{\em survives} in the forward limit,\cite{diehl}
namely the tensor charge.

Additionally, it has been pointed out by Diehl, that angular 
momentum conservation in these helicity flip
amplitudes is accompanied  by a transfer of 
orbital angular momentum.  This is
indicated by nonzero intrinsic transverse momentum transfer  of the
partons which is not observed in ordinary parton distribution 
functions where the incoming and outgoing nucleon momentum are equal. 
We find this essential  correlation between helicity flip, orbital angular
momentum and dependence on transverse momentum transfer
to persist in our model estimate of the tensor charge.

\section{Modeling The Tensor Charge }

Here we present an approach to calculating the tensor charge 
that exploits the  approximate mass degeneracy of the light
axial vector mesons ($a_1$(1260), $b_1$(1235) and $h_1$(1170)) and
uses pole dominance to calculate the tensor  
charge.\cite{gamgold1,gamgold2}
Our motivation stems in part from the observation
that the tensor charge does not mix with gluons under QCD
evolution and therefore behaves as a non-singlet
matrix element.   In conjunction with the fact that the
tensor current is charge conjugation odd (it does not mix
quark-antiquark excitations of the vacuum, since the latter is charge
conjugation even) suggests that the 
tensor charge is amenable to a  valence quark
model analysis.

\subsection{Pole Dominance and Spin-Flavor Symmetry}

The flavor components of the nucleon tensor charge are defined 
from the local operator nucleon matrix element of the tensor current,
\begin{equation}
\langle P,S_{T}\big|\overline{\psi}
\sigma^{\mu\nu}\gamma_5 \frac{\lambda^a}{2}\psi\big| P,S_{T}\rangle
\hspace{-.05cm}=\hspace{-.05cm}2\delta
q^a(\mu^2)(P^{\mu}S^{\nu}_T\hspace{-.05cm}-\hspace{-.05cm}P^{\nu}S^{\mu}_T).
\label{eq1}
\end{equation}
We adopt the model that the nucleon matrix element of the 
tensor current is dominated by the lowest lying axial
vector mesons
\be
\langle P,S_{T}\Big|\overline{\psi}
\sigma^{\mu\nu}\gamma_5 \frac{\lambda^a}{2}\psi\Big| P,S_{T}\rangle=
\lim_{k^2\rightarrow 0}\sum_{\M} \frac{\langle 0\big|
\overline{\psi}
\sigma^{\mu\nu}\gamma_5 \frac{\lambda^a}{2}
\psi
\big|\M\rangle\langle \M , P,S_{T}| P,S_{T}
\rangle}{M^2_{\M}-k^2} .
\label{eq4}
\ee
The summation is over those mesons with quantum numbers,
$J^{PC}=1^{+-}$ that  couple to the nucleon via the tensor current;
namely  the charge conjugation odd axial vector mesons -- the isoscalar
$h_1(1170)$ and the isovector $b_1(1235)$.
To analyze this expression in the limit $k^2\rightarrow 0$
we require the vertex functions for the nucleon coupling to the
$h_1$ and $b_1$ meson 
\be
\langle M P| P\rangle=
\frac{i\gmnn}{2M_N}\overline{u}\left({\scriptstyle{P,S_{T}}}\right)
\sigma^{\mu\nu}\gamma_5
u\left({\scriptstyle{P,S_{T}}}\right)\eps_\mu k_\nu ,
\ee
and the corresponding matrix elements
of the meson decay amplitudes which are related to the meson to vacuum
matrix element via the quark tensor current.
\be
\langle 0\Big|
\overline{\psi}
\sigma^{\mu\nu}\gamma_5 \frac{\lambda^a}{2}\psi
\Big|\M\rangle=
if^a_{\M}\left(\eps_\mu k_\nu-\eps_\nu k_\mu\right) .
\ee
Here $P_\mu$ is the nucleon  momentum,
and  $k_\mu$ and $\eps_\nu$ are the meson momentum and
polarization respectively.
The former yield the nucleon coupling constants
$\gmnn$  and the latter yield the meson decay constant $f_{\M}$.
Taking a  hint from the valence interpretation of the tensor
charge, we  exploit the phenomenological mass
symmetry among the lowest lying axial vector mesons that 
couple to the tensor charge; we adopt the spin-flavor 
symmetry characterized by an $SU(6)_W \otimes O(3)$ multiplet 
structure.\cite{sakita}   Thus, the $1^{+-}$ $h_1$ and $b_1$ mesons fall into a
$\left(35\otimes L=1\right)$ multiplet that contains
$J^{PC}=1^{+-},0^{++},1^{++},2^{++}$ states, where 
these mesons couple ``symmetrically'' to  baryons 
\be
\Tr(J\cdot\Phi)=g\left(\quad .\, .\, .\, +\, 
c_1\frac{J_{\mu\nu}^{5\, a}F^{\mu\nu}_a}{4M_N}+\, 
c_2 J_\mu^{5\, a}A_{ a}^\mu\, +\quad .\, .\, .\, \right) ,
\nnn
\label{six}
\ee
where $J$ and $ \Phi$ are the nucleon ``super'' current 
and meson ``super''multiplet.
Reducing  this expression 
to 2-component form 
\be
{\cal L}^{\scriptstyle{(SU(6)\otimes O(3))}}_{\M NN}=g\ N^{\da}\left(\quad .\, .\, .\, +\, 
\frac{5}{3}\sigma\cdot \hk  \hP\cdot\eps_{\bia}
+\frac{i}{\sqrt{2}}\left(\hP\times \hk\right)\cdot \eps_{\ai}
+\quad .\, .\, .\,
\right)N ,
\nnn
\label{seven}
\ee 
we identify the  $SU(6)_W\otimes O(3)$ Yukawa
couplings, the $c'^{\rm s}$, and thus the $\gmnn$.  
Similarly,  the meson decay constants are determined from
the $SU(6)_W\otimes O(3)$ {\em quark current} couplings to the mesons,
\be
{\cal L}^{\scriptstyle{(SU(6)\otimes O(3))}}_{\M qq}=f\ \chi^{\da}\left(\quad .\, .\, .\, +\, 
\sigma\cdot \hk \hP\cdot \eps_{\bia}
+\frac{i}{\sqrt{2}}\left(\hP\times \hk\right)\cdot \eps_{\ai}
+\quad .\, .\, .\, \right)\chi .
\label{quark}
\nnn
\ee
This analysis enables us to relate the $a_1$ meson
decay constant measured in $\tau$ 
decay,~\cite{tsai} $f_{\ai} =(0.19\pm 0.03) {\rm GeV^2}$,
and the  $a_1 N N$ coupling constant $g_{\aiNN}=7.49\pm 1.0$
(as determined from $a_1$ axial vector dominance for longitudinal
 charge as derived in Ref.\cite{birkel} but using
$g_A/ g_V= 1.267$~\cite{pdg})
to the meson decay constants and coupling
constants. We find 
\be
f_{\bia}= \frac{\sqrt{2}}{M_{\bia}}f_{\ai}, \quad
g_{\biNN}=\frac{5}{3 \sqrt{2}} g_{\aiNN},
\label{ok1}
\ee
where the $5/ 3$ appears from the $SU(6)$ factor $(1+F/ D)$
and the $\sqrt{2}$ arises from the $L=1$ relation between
the $1^{++}$ and $1^{+-}$ states. Our resulting value
of $f_{\bia}\approx 0.21\pm 0.03$ agrees well with a
sum rule determination of $0.18\pm 0.03$.\protect\cite{ball}  The 
$h_1$ couplings are related to the $b_1$ couplings via $SU(3)$
and the $SU(6)$ $F/D$ value,
\be
f_{\bia}=\sqrt{3}f_{\hi}, \quad 
g_{\biNN}=\frac{5}{\sqrt{3}}g_{\hiNN}
\label{ok2}
\ee
For transverse polarized Dirac particles, $S^\mu=(0,S_T)$ these 
values, in turn,  enable us to determine the isovector and isoscalar parts
of the tensor charge,
\be
\delv =\frac{f_{\bia}g_{\biNN}\langle k_{\perp}^2\rangle }{\sqrt{2} M_N
M_{\bia}^2}\, ,\quad
\dels = \frac{ f_{\hi}g_{\hiNN}\langle k_{\perp}^2\rangle }{\sqrt{2} M_N
M_{\hi}^2},
\ee
respectively (where,
$\delv\hspace{-.1cm}=\hspace{-.1cm}(\delta u-\delta d)$, and
$\dels\hspace{-.1cm}=\hspace{-.1cm}(\delta u+\delta d)$).
Transverse momentum appears in these expressions because the tensor
couplings involve helicity flips that carry kinematic factors of
3-momentum transfer, as required by rotational invariance. The squared
4-momentum transfer of the external hadrons goes to zero in
Eq.~(\ref{eq4}), but the quark fields carry intrinsic transverse
momentum. This intrinsic $k_{\perp}$ of the quarks
in the nucleon is determined from Drell-Yan processes and from heavy
vector boson production. 
We use a Gaussian momentum distribution, and
let $\langle k_{\perp}^2\rangle$ range from
$\left(0.58 \:{\rm to}\: 1.0\   {\rm GeV}^2\right)$.\cite{ellis}

\section{Mixing and Results}

In relating the $b_1(1235)$ and $h_1(1170)$ couplings in Eq.~(\ref{ok2}) 
we assumed that the latter isoscalar was a pure octet element,
$h_1(8)$.  Experimentally,  the higher mass $h_1(1380)$
was seen in the {}$K+\bar{K}+\pi's$ decay
channel~\cite{pdg,abele} while the $h_1(1170)$ was detected in the
multi-pion
channel.\cite{pdg,ando}   This decay pattern indicates that the higher mass
state is strangeonium and decouples from the lighter quarks -- the well
known mixing pattern of the vector meson nonet elements $\omega$ and
$\phi$. If the $h_1$ states are mixed states of the $SU(3)$ octet $h_1(8)$
and singlet $h_1(1)$ analogously, then
it follows that
\be
f_{\hit} = f_{\bia}\, ,  \quad   g_{\hitNN} =
\frac{3}{5}g_{\biNN},
\label{mixing}
\ee
with the $h_1(1380)$ not coupling to the nucleon (for
$g_{\hi(1)NN}=\sqrt{2}g_{\hi(8)NN}$). These symmetry relations
yield the results 
\be
\delta u(\mu^2)=(0.58 \:{\rm to}\: 1.01)\pm 0.20, \quad
\delta d(\mu^2)=-(0.11 \:{\rm to}\: 0.20)\pm 0.20.
\label{newcharges}
\ee
These values are similar to several other 
model calculations: from the lattice; to
QCD sum rules; the bag model; and quark soliton 
models\cite.{jihe}  The calculation has been carried out at 
the scale $\mu\approx 1$ GeV,
which is set by the nucleon mass as well as being the mean mass of the
axial vector meson multiplet. The appropriate evolution to higher scales
(wherein the Drell-Yan processes are studied) is determined by the
anomalous dimensions of the tensor charge\cite{artru}
which is straightforward but a slowly varying.

It is interesting to observe that the symmetry relations that connect
the $b_1$ couplings to the $a_1$ couplings in Eq.~(\ref{ok1}) can be
used to relate directly the isovector tensor charge to the axial vector
coupling $g_A$. This is accomplished through the $a_1$ dominance
expression for the isovector longitudinal charges derived in,\cite{birkel}
\be
\Delta u - \Delta d = \frac{g_A}{g_V}=
\frac{\sqrt{2} f_{\ai} g_{\aiNN}}{M_{\ai}^2}.
\label{eqbirkel}
\ee
Hence for $\delv$ we have
\be
\delta u -\delta d =\frac{5}{6}\frac{g_A}{g_V}\frac{ M_{\ai}^2}{
M_{\bia}^2}\frac{\langle k_{\perp}^2\rangle}{ M_N M_{\bia}} \, ,
\label{eqgtga}
\ee
It is important to realize that this relation can hold at the
scale wherein the couplings were specified, the meson masses, but will be
altered at higher scales (logarithmically) by the different evolution
equations for the $\Delta q$ and $\delta q$ charges. To write an analogous
expression for the isoscalar charges
($\Delta u + \Delta d$) would involve the singlet mixing terms and gluon
contributions, as Ref.\cite{birkel} considers. However, given that the
tensor charge does not involve gluon contributions (and anomalies), it
is expected that the relation between the $h_1$ and $b_1$ couplings
in the same $SU(3)$ multiplet will lead to a more direct result
\be
\delta u+\delta d = \frac{3}{5}\frac{M_{\bia}^2}{M_{\hi}^2}\delv\, ,
\ee
for the ideally mixed singlet-octet $h_1(1170)$. These 
relations  are quite distinct from other predictions.

\section{Conclusions}
Our axial vector dominance model with
$SU(6)_W \otimes O(3)$ coupling relations provide simple formulae for the
tensor charges. This simplicity obscures the considerable subtlety of the
(non-perturbative) hadronic physics that is summarized in those
formulae. 
These results support the view that the underlying
hadronic physics, while quite difficult to formulate from first
principles, is essentially a $1^{+-}$ meson exchange 
process.  Further, through the kinematics of DVCS and those indicated by
low momentum transfer we find the essential correlation between 
helicity flip, orbital angular
momentum and dependence on transverse momentum transfer
to persist in our model estimate of the tensor charge.

\section*{Acknowledgments} L.G. 
is grateful to the organizers of the 
 $3^{\rm rd}$ Circum Pan-Pacific Symposium on High Energy Spin Physics, 
for their hospitality and 
and in particular Bo-Qiang Ma for his
tireless organizational  work during and after 
the conference. Also, I thank Shunzo Kumano  for the original 
invitation to  the conference.
Finally, I thank Stan Brodsky, Markus Diehl, 
Robert  Jaffe, Xiangdong Ji and Wally Melnitchouk
for valuable comments on this work.
This work is supported in part by Grant No. 
DE-FG02-92ER40702 from the U.S. Department of Energy.


\begin{thebibliography}{0}

\bibitem{smc} J. Ashman {\em et al.} Nucl. Phys. {\bf B328}, 1 (1989);
B. Adeva {\em et al.}, Phys. Lett. B {\bf 302}, 533 (1993);
P.L. Anthony {\em et al.}, Phys. Rev. Lett. {\bf 71}, 959 (1993) ;
K. Abe {\em et al.}, Phys. Rev. Lett. {\bf 72}, 25 (1995).
\bibitem{ji_spin} X. Ji, Phys. Rev. Lett {\bf 78}, 610 (1997).
\bibitem{review} V. Barone ,  A. Drago and P. G. Ratcliffe,
arXive: hep-ph/0104283 and references contained therein .

\bibitem{transversity}  G.R. Goldstein and M.J. Moravcsik,
  Ann. Phys. (NY) {\bf 98}, 128 (1976); Ann. Phys. (NY) {\bf 142}, 219
  (1982); Ann. Phys. (NY){\bf 195}, 213 (1989).

\bibitem{jaffe91}
R.\ L.\ Jaffe and X.\ Ji, Phys. Rev. Lett. {\bf 67}, 552 (1991);
Nucl. Phys. {\bf B375}, 527 (1992).

\bibitem{soffer95}
J.\ Soffer, Phys. Rev. Lett. {\bf 74}, 1292 (1995); G.\ R. \ Goldstein,
R.\ L.\ Jaffe and X.\ Ji, Phys. Rev. D {\bf 52}, 5006 (1995).

\bibitem{jihe}
H.  He and X. Ji, Phys. Rev. D {\bf 52}, 2960 (1995);
Phys. Rev. D {\bf 54}, 6897 (1996);
S. Aoki, {\em et al.}, Phys. Rev. D {\bf 56}, 433 (1997);
L. Gamberg, H. Reinhardt and H. Weigel, 
Phys. Rev. D {\bf 58}, 054014 (1998); L. Gamberg,
``Structure Functions and Chiral Odd Quark Distributions in
the NJL Soliton Model of the Nucleon'',  Proceedings of
Future Transversity Measurements, RIKEN-BNL Workshop (2000).

\bibitem{rad} X. Ji, Phys. Rev. Lett {\bf 78}, 610 (1997);
A. V. Radyushkin, Phys. Rev. D{\bf 56}, 5524 (1997).

\bibitem{diehl} M. Diehl, Eur. Phys. J. C{\bf 19}, 485 (2001).

\bibitem{gamgold1} L. Gamberg and G. R. Goldstein,
arXive: hep-ph/0106178.
\bibitem{gamgold2} L. Gamberg and G. R. Goldstein,
Phys. Rev. Lett. {\bf 87}, 242001 (2001).

\bibitem{sakita} B. Sakita, Phys. Rev. {\bf 136}, B1756
(1964); F. G\"{u}rsey and L.A. Radicati, Phys. Rev. Lett. {\bf 13}, 173
(1964).

\bibitem{tsai} W.-S. Tsai, Phys. Rev. D {\bf 4}, 2821 (1971).

\bibitem{birkel} M. Birkel and H. Fritzsch, Phys. Rev. D {\bf 53}, 6195
(1996).
\bibitem{pdg} D.E. Groom, {\it et al.} (Particle Data Group),
Eur. Phys. Jour. {\bf C15}, 1 (2000).
\bibitem{ball}
V.M. Belyaev and A. Oganesian , Phys. Lett. {\bf B395},
307 (1997).
\bibitem{ellis} R.K. Ellis, W.J. Stirling and B.R. Webber, {\it QCD and
Collider Physics} (Cambridge University Press, Cambridge, U.K. 1996),
p.305.
\bibitem{abele} A. Abele, {\it et al.}, Phys. Lett. {\bf B415}, 280
(1997); D. Aston, {\it et al.}, Phys. Lett. {\bf B201}, 573 (1988).
\bibitem{ando} A. Ando, {\it et al.}, Phys. Lett. {\bf B291}, 496
(1992); and references contained therein.

\bibitem{artru}
X.\ Artu and M.\ Mekhfi, Z. Phys. {\bf C45} (1990) 669.


\end{thebibliography}
\end{document}